\documentclass[prc,showpacs,preprint,superscriptaddress,nofootinbib]{revtex4}%
\usepackage{epsfig,float}
\usepackage{amsmath}
\usepackage{amsfonts}
\usepackage{amssymb}
\usepackage{graphicx}%
\begin{document}
\title{An explanation of the $\Delta_{5/2^{-}}(1930)$ as a $\rho\Delta$ bound state.}
\author{P. Gonz\'{a}lez}
\affiliation{Departamento de F\'{\i}sica Te\'{o}rica, Universidad de Valencia (UV) and IFIC
(UV-CSIC), Institutos de Investigaci\'{o}n de Paterna, Aptdo. 22085, 46071
Valencia, Spain}
\author{E. Oset}
\affiliation{Departamento de F\'{\i}sica Te\'{o}rica, Universidad de Valencia (UV) and IFIC
(UV-CSIC), Institutos de Investigaci\'{o}n de Paterna, Aptdo. 22085, 46071
Valencia, Spain}
\author{J. Vijande}
\affiliation{Departamento de F\'{\i}sica At\'{o}mica, Molecular y Nuclear, Universidad de
Valencia (UV) and IFIC (UV-CSIC), Valencia, Spain.}
\date{today }

\begin{abstract}
We use the $\rho\Delta$ interaction in the hidden gauge formalism to
dynamically generate $N^{\ast}$ and $\Delta^{\ast}$ resonances. We show,
through a comparison of the results from this analysis and from a quark model
study with data, that the $\Delta_{5/2^{-}}(1930),$ $\Delta_{3/2^{-}}(1940)$
and $\Delta_{1/2^{-}}(1900)$ resonances can be assigned to $\rho\Delta$ bound
states. More precisely the $\Delta_{5/2^{-}}(1930)$ can be interpreted as a
$\rho\Delta$ bound state whereas the $\Delta_{3/2^{-}}(1940)$ and
$\Delta_{1/2^{-}}(1900)$ may contain an important $\rho\Delta$ component. This
interpretation allows for a solution of a long-standing puzzle concerning the
description of these resonances in constituent quark models. In addition we
also obtain degenerate $J^{P}=1/2^{-},3/2^{-},5/2^{-}$ $N^{*}$ states but
their assignment to experimental resonances is more uncertain.

\end{abstract}

\maketitle

\section{Introduction}

\label{s1}

The interpretation of spectra of baryons is still a thriving field that is
attracting much attention. The traditional view of baryons as made of three
constituent quarks~\cite{quarks} is being substituted by a more extended view
of baryonic states involving three quark $(3q)$ as well as four quark-one
antiquark $(4q1\overline{q})$ components. In particular some baryonic
resonances (a paradigmatic case is the $\Lambda(1405))$ may be better
interpreted as molecular states of mesons and baryons. Though such ideas have
been advocated in the past \cite{Dalitz:1959dn}, it has only been in recent
years that detailed quantitative studies have been done based on the
combination of chiral dynamics with unitary nonperturbative techniques in
coupled channels of mesons and baryons. Thus the low lying $J^{P}=1/2^{-}$
baryonic resonances are relatively well interpreted in terms of meson--baryon
molecules~\cite{weise,ollerulf}, indicating the relevance of these
$4q1\overline{q}$ components for their description. More technically, they are
dynamically generated from the interaction of the octet of mesons containing
the $\pi$ and the octet of baryons including the proton. Similarly, the
interaction of the octet of mesons of the $\pi$ with the baryon decuplet of
the $\Delta$ leads to dynamically generated states of $J^{P}=3/2^{-}$, which
can be associated to existing resonances~\cite{kolodecu,sarkar}. A further step in
this molecular engineering is done with the study of $J^{P}=1/2^{+}$ states
stemming from the interaction of a pair of pseudoscalar mesons with a baryon
of the octet of the nucleon \cite{alberto}, corresponding indeed to
$5q2\overline{q}$ components. A common denominator of these generated baryon
states is the use of pseudoscalar mesons as building blocks. Here we undertake
the task of extending the study of the dynamical generation of resonances to
the vector meson-baryon sector. Besides having its own interest as an
extension of the theoretical formalism, this study is particularly relevant in
this moment from a phenomenological point of view since there are clear
indications that $\rho N$ components \cite{Zou08} as well as $\rho\Delta$
($\omega\Delta)$ ones \cite{GVV08} may play an essential role in the precise
description of the negative parity $\Delta$ spectrum below $2.0$ GeV.

A framework that makes the study of vector mesons interacting with baryons
accurate and manageable is the hidden gauge formalism \cite{hidden1}. There,
pseudoscalars and vectors are introduced with an interaction which respects
chiral symmetry. The consideration of the interaction of vector mesons with
baryons allows then to reinterpret the pseudoscalar meson-baryon chiral
Lagrangians as the result of the exchange of vector mesons in the t-channel.
The novelty of such a framework is that it also contains the coupling of
vector mesons among themselves and therefore one can construct their
interaction with baryons. The combination of such an interaction with chiral
unitary techniques has been rather successful. In particular, the $\rho\rho$
interaction has been shown to lead to the dynamical generation of the
$f_{2}(1270)$ and $f_{0}(1370)$ resonances \cite{raquel}, with a branching
ratio for the sensitive $\gamma\gamma$ decay channel in good agreement with
experimental data~\cite{junko}.

In this work we present the formalism and results for the $\rho\Delta
\rightarrow\rho\Delta$ interaction which leads to baryonic states in fair
agreement with known resonances, within experimental and theoretical
uncertainties. The approximate degeneracy of the experimental $J^{P}%
=1/2^{-},3/2^{-},5/2^{-}$ $\Delta^{\ast}$ states around $1920$ MeV appears as
a dynamical feature of the theory. In the case of the predicted $J^{P}%
=1/2^{-},3/2^{-},5/2^{-}$ $N^{\ast}$ states an assignment to known $N^{\ast}$
resonances around $1700$ MeV is also feasible although the presence of
corresponding $3q$ states close below in mass points out to the need of
incorporating both components ($3q$ and $\rho\Delta)$ in their description.
These contents are distributed as follows. In section \ref{s2} the formalism
for the dynamical generation of resonances from the $\rho\Delta\rightarrow
\rho\Delta$ interaction is derived. In Sections \ref{s3}, \ref{s4} and
\ref{s5} the results obtained in different approximations are presented.
Section \ref{s6} is devoted to an analysis of the possible contribution from
anomalous terms involving $\rho\omega\pi$ vertexes. The comparison of our
results with experimental data is done in section \ref{s7}. Finally in section
\ref{s8} we establish our main conclusions.

\section{Formalism for the $VV$ and $VB$ interaction}

\label{s2} 
\begin{figure}[tb]
\begin{center}
\epsfig{file=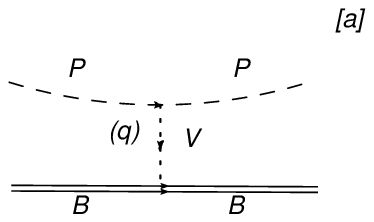, width=7cm} \epsfig{file=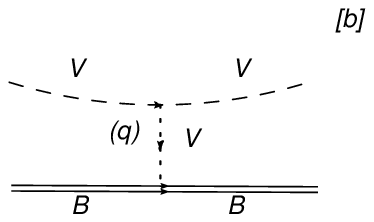, width=7cm}
\end{center}
\caption{Diagrams obtained in the effective chiral Lagrangians for interaction
of pseudoscalar [a] or vector [b] mesons with the decuplet of baryons.}%
\label{f1}%
\end{figure}

We follow the formalism of the hidden gauge interaction for vector
mesons \cite{hidden1} (see also~\cite{hidekoroca} for a practical set of
Feynman rules). The Lagrangian involving the interaction of vector mesons
among themselves is given by
\begin{equation}
\mathcal{L}_{III}=-\frac{1}{4}\langle V_{\mu\nu}V^{\mu\nu}\rangle\ ,
\label{lVV}%
\end{equation}
where the symbol $\langle\rangle$ stands for the trace in the $SU(3)$ space
and $V_{\mu\nu}$ is expressed as
\begin{equation}
V_{\mu\nu}=\partial_{\mu}V_{\nu}-\partial_{\nu}V_{\mu}-ig[V_{\mu},V_{\nu}]\ ,
\label{Vmunu}%
\end{equation}
with $g$ given by
\begin{equation}
g=\frac{M_{V}}{2f}\ , \label{g}%
\end{equation}
with $f=93$ MeV being the pion decay constant. The value of $g$ of Eq.
(\ref{g}) is one of the ways to account for the $KSFR$ rule~\cite{KSFR} which
is tied to vector meson dominance~\cite{sakurai}. The magnitude $V_{\mu}$ is
the $SU(3)$ matrix of the vectors of the octet of the $\rho$
\begin{equation}
V_{\mu}=\left(
\begin{array}
[c]{ccc}%
\frac{\rho^{0}}{\sqrt{2}}+\frac{\omega}{\sqrt{2}} & \rho^{+} & K^{\ast+}\\
\rho^{-} & -\frac{\rho^{0}}{\sqrt{2}}+\frac{\omega}{\sqrt{2}} & K^{\ast0}\\
K^{\ast-} & \bar{K}^{\ast0} & \phi\\
&  &
\end{array}
\right)  _{\mu}\ . \label{Vmu}%
\end{equation}
The interaction of $\mathcal{L}_{III}$ gives rise to a contact term coming for
$[V_{\mu},V_{\nu}][V^{\mu},V^{\nu}]$ of the form
\begin{equation}
\mathcal{L}_{III}^{(c)}=\frac{g^{2}}{2}\langle V_{\mu}V_{\nu}V^{\mu}V^{\nu
}-V_{\nu}V_{\mu}V^{\mu}V^{\nu}\rangle\ , \label{lcont}%
\end{equation}
and also to a three vector vertex,
\begin{equation}
\mathcal{L}_{III}^{(3V)}=ig\langle(\partial_{\mu}V_{\nu}-\partial_{\nu}V_{\mu
})V^{\mu}V^{\nu}\rangle\ . \label{l3V}%
\end{equation}
It is useful to rewrite this last term as:
\begin{align}
\label{10}\mathcal{L}_{III}^{(3V)}  &  =ig\langle V^{\nu}\partial_{\mu}%
V_{\nu}V^{\mu}-\partial_{\nu}V_{\mu}V^{\mu}V^{\nu}\rangle=ig\langle V^{\mu
}\partial_{\nu}V_{\mu}V^{\nu}-\partial_{\nu}V_{\mu}V^{\mu}V^{\nu}%
\rangle\\
&  =ig\langle\left(  V^{\mu}\partial_{\nu}V_{\mu}-\partial_{\nu}V_{\mu}V^{\mu
}\right)  V^{\nu}\rangle\nonumber
\end{align}
in complete analogy with the coupling of a vector to pseudoscalar mesons in
the same theory, which is given in Ref.~\cite{hidden1} as
\begin{equation}
\mathcal{L}^{VPP}=-ig\langle\left[  \Phi,\partial_{\nu}\Phi\right]  V^{\nu
}\rangle\label{11}%
\end{equation}
with $\Phi$, the analogous matrix to Eq. (\ref{Vmu}), containing the
pseudoscalar fields ($P$). This analogy allows us to obtain the interaction of
vector mesons with the decuplet of baryons in a straightforward way by
realizing that the chiral Lagrangian of Ref.~\cite{manohar} for the
interaction of the octect of pseudoscalar mesons with the decuplet of baryons
is obtained from the exchange of a vector meson between the pseudoscalar
mesons and the baryon, as depicted in Fig.\ref{f1}[a], in the limit of
$q^{2}/M_{V}^{2}\rightarrow0$, being $q$ the momentum transfer.

Then by substituting the vertex of Eq.~(\ref{11}) by that of Eq.~(\ref{10})
the physical picture goes from diagram [a] to [b] of Fig.\ref{f1}. Notice that
diagram [b] has a more complicated structure since one has three vector
fields to destroy or create either vector, whereas in diagram [a] there is only
one vector field and hence no choice. Yet, in the implicit approximations
leading to the effective chiral Lagrangian one can go a step further, in line
with neglecting $q^{2}/M_{V}^{2}$. Thus, we will neglect the three momentum of
the external vectors relative to their mass. Since the vector polarizations
have $\epsilon^{0}$ either zero for the transverse polarizations, or 
$|\vec{k}|/M_{V}$ , with $\vec{k}$ the vector meson
trimomentum, for the longitudinal component, all the external vectors will
have zero $\epsilon^{0}$ component. Following the same argument, $V^{\nu}$,
appearing in the term $\partial_{\nu}V^{\nu}$ in Eq.~(\ref{10}), cannot be an
external vector since then $\partial_{\nu}$ would lead to a three-momentum,
which is neglected in the approach. Then $V^{\nu}$ corresponds to the
exchanged vector, $V^{\mu}V_{\mu}$ gives rise to $-\vec{\epsilon}%
\,\vec{\epsilon}^{\text{ }\prime}$ for the external vectors and then the PB
and VB interactions become formally identical (including the sign), with
$\vec{\epsilon}\,\vec{\epsilon}^{\text{ }\prime}$ factorized in the VB
interaction. For practical reasons one can replace the matrix $\Phi$ by the
matrix $V$ in the interaction Lagrangian of Ref.~\cite{manohar}, and this is
equivalent to substituting $\pi^{+}\rightarrow\rho^{+}$, $K^{+}\rightarrow
K^{\ast+}$, etc... in the matrix elements of PB$\rightarrow$ PB to obtain
those of VB$\rightarrow$VB. There is only a small amendment to be done
concerning the $\phi$ and $\omega$ since the ideal mixing implies:
\begin{align}
\omega &  =\frac{2}{\sqrt{6}}\omega_{1}+\frac{1}{\sqrt{3}}\omega_{8}\\
\phi &  =\frac{1}{\sqrt{3}}\omega_{1}-\frac{2}{\sqrt{6}}\omega_{8}%
\nonumber\label{12}%
\end{align}
and the structure of Eq.~(\ref{10}) does not give any contribution for the
$\omega_{1}$, which appears diagonal as diag($\frac{1}{\sqrt{3}}\omega
_{1},\frac{1}{\sqrt{3}}\omega_{1},\frac{1}{\sqrt{3}}\omega_{1}$) in the $V$
matrix. As a consequence of that, the matrix elements of the potential
involving $\omega(\phi)$ are obtained from those of $\eta_{8}$ of
$PB\rightarrow PB$ multiplying by $\frac{1}{\sqrt{3}}(-\sqrt{\frac{2}{3}})$.

After this discussion we can use directly the results of Ref.~\cite{sarkar} to
get the $\rho\Delta\rightarrow\rho\Delta$ potential with different charges. We
find
\begin{equation}
V_{ij}=-\frac{1}{4f^{2}}C_{ij}(k^{0}+k^{^{\prime}0})\vec{\epsilon}%
\,\vec{\epsilon}^{\text{ }\prime}\label{13}%
\end{equation}
where $k^{0}$ and $k^{^{\prime}0}$ are the energies of the incoming/outgoing
vectors and the $C_{ij}$ coefficients are given in Ref.~\cite{sarkar}
substituting $\pi$ by $\rho$. One can do the corresponding isospin projections
and the results are equally given in Ref.~\cite{sarkar} (for $I=5/2$,
corresponding to $\Delta^{++}\pi^{+}\rightarrow\Delta^{++}\pi^{+}$, the result
can be found in the appendix of this reference):
\begin{align}
\rho\Delta\,,I &  =\frac{1}{2}\,:\,C=5\label{ces}\\
\rho\Delta\,,I &  =\frac{3}{2}\,:\,C=2\nonumber\\
\rho\Delta\,,I &  =\frac{5}{2}\,:\,C=-3\nonumber
\end{align}

As in Ref.~\cite{sarkar} we shall solve the Bethe-Salpeter equation with the
on-shell factorized potential~\cite{nsd,ollerulf} and, thus, the T--matrix
will be given by
\begin{equation}
T=\frac{V}{1-VG}\vec{\epsilon}\,\vec{\epsilon}^{\text{ }\prime}\,,\label{15}%
\end{equation}
being $V$ the potential of Eq.~(\ref{13}) in the isospin basis removing the
factor $\vec{\epsilon}\,\vec{\epsilon}^{\text{ }\prime}$, and $G$ the loop
function for intermediate $\rho\Delta$ states, also given in
Ref.~\cite{sarkar}, regularized both with a cuttoff prescription or with
dimensional regularization.

In the present work we ignore the coupling with the $\Sigma^{\ast}K^{\ast} $
channel. The reason being that having a mass 275 MeV above the $\rho\Delta$
threshold it is expected not to play a relevant role in the description of the
$\rho\Delta$ states.

The prescription after Eq.(\ref{13}) to obtain couplings to $\rho
\Delta\rightarrow\omega\Delta$ or $\omega\Delta\rightarrow\omega\Delta$,
together with the tables of Ref.~\cite{sarkar}, gives zero for these
transitions. Actually, this is easy to visualize in the vector exchange model
since the $\rho\rho\omega$ and $\omega\omega\omega$ vertexes violate
$G-$parity while the $\rho\omega\omega$ vertex violates isospin. Thus, for the
problem of the lightest states of the vector-baryon decuplet one can consider
single $\rho\Delta$ channels.

\begin{figure}[tb]
\begin{center}
\vspace*{1cm} \epsfig{file=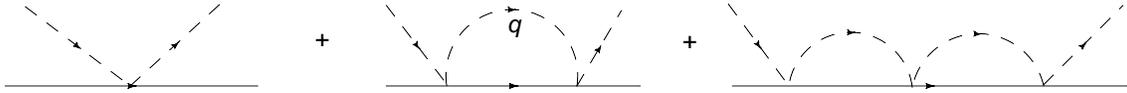, width=15cm}
\end{center}
\caption{Terms appearing in the Bethe--Salpeter equation of Eq.(\ref{15}).}%
\label{f2}%
\end{figure}

For the interactions implicit in the Bethe--Salpeter equation, see
Fig.~\ref{f2}, one has at second order
\begin{equation}
-it=-iV\epsilon_{i}\epsilon_{i}iG(-i)V\epsilon_{j}\epsilon_{j}\label{16}%
\end{equation}
with $i,j$ spatial components, and summing over polarizations
\begin{equation}
\sum_{\lambda}\epsilon_{i}\epsilon_{j}=\delta_{ij}+\frac{q_{i}q_{j}}{m_{V}%
^{2}}\,.
\end{equation}
In Ref.~\cite{luisaxial} following the on-shell factorization, the
$\frac{q_{i}q_{j}}{m_{V}^{2}}$ term was included in the loop function, giving
rise to a correction term $\frac{\vec{q}^{\,\,2}}{3m_{V}^{2}}$ which was very
small. Consistently with the approximations done here , $\frac{q^{2}}%
{m_{V}^{2}}=0$, we also neglect this term. The factor $\vec{\epsilon}%
\,\vec{\epsilon}^{\text{ }\prime}$ appears in all iterations and, thus,
factorizes in the T--matrix.

\section{Results with no width for $\rho$ and $\Delta$.}

\label{s3} From Eqs.~(\ref{13}), (\ref{ces}) and (\ref{15}) we find two
attractive, $I=1/2$ and $I=3/2$, and one repulsive, $I=5/2$, channels. Hence,
no bound states are obtained in the exotic $I=5/2$ channel. On the contrary,
bound states ($|T|^{2}$ goes to infinity in Eq.~(\ref{15})) can be clearly
observed in the $I=1/2$, and $I=3/2$ channels. The strength of the interaction
indicates that the $I=1/2$ state is more bound than the $I=3/2$.

Another issue worth mentioning is that the only spin dependence comes from the
$\vec{\epsilon}\,\vec{\epsilon}^{\text{ }\prime}$ factor of the vector mesons.
The spin of the $\Delta$ does not appear in our formalism due to the
approximations done. Only the $\gamma^{0}$ term of the $VBB$ vertex, which has
not spin structure to leading order, has been kept~\cite{sarkar}. The
$\vec{\epsilon}\,\vec{\epsilon}^{\text{ }\prime}$ scalar structure tells us
that all spin states of the $\rho\Delta$ system behave according to the same
interactions and therefore, one has degeneracy for the $J^{P}=1/2^{-}%
,3/2^{-},5/2^{-}$ states, both for $I=1/2$ and $I=3/2$.

\begin{figure}[tb]
\begin{center}
\vspace*{-2cm} \epsfig{file=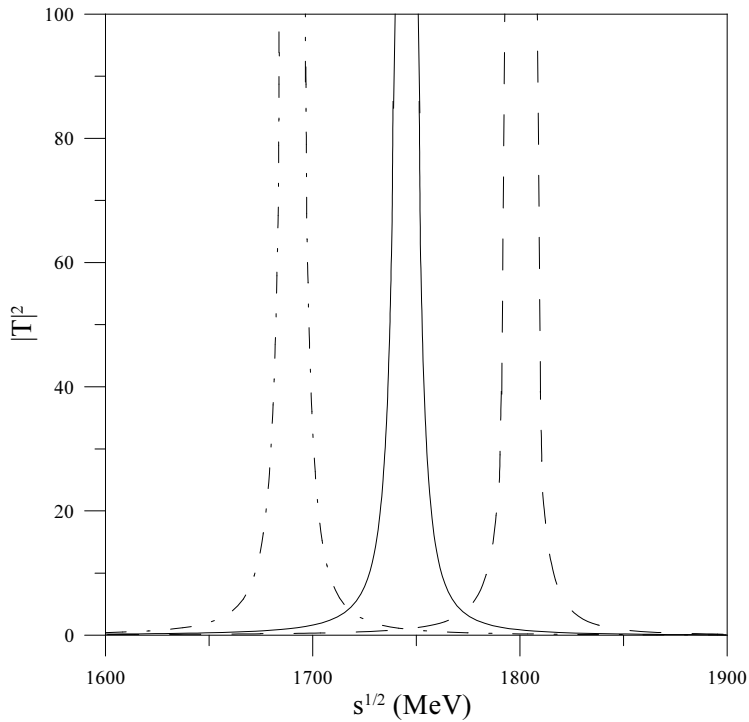, width=10cm} \vspace*{-3cm}
\end{center}
\caption{$|T|^{2}$ for $\rho\Delta\rightarrow\rho\Delta$ in the $I=1/2$
channel for several values of the cutoff $q_{max}$: solid line $q_{max}=770$
MeV, dashed line $q_{max}=700$ MeV, dashed-dotted line $q_{max}=630$ MeV.}%
\label{f3}%
\end{figure}

\begin{figure}[tb]
\begin{center}
\vspace*{-2cm} \epsfig{file=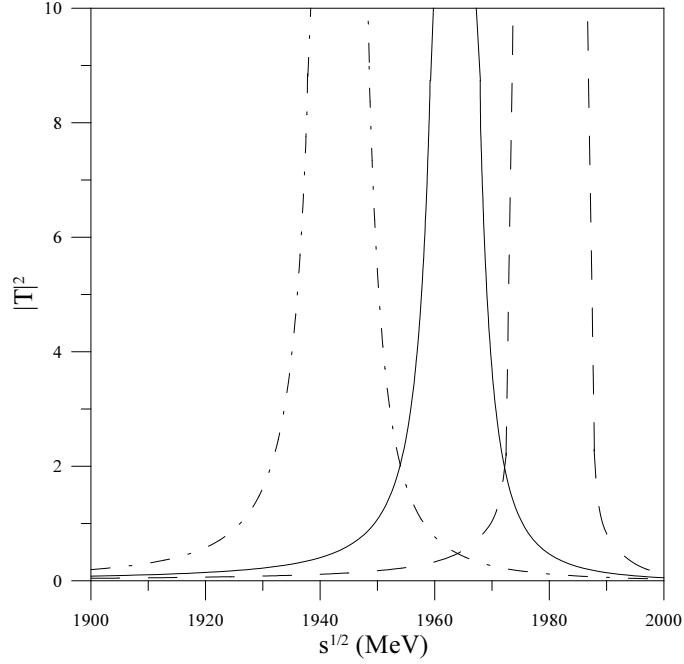, width=10cm} \vspace*{-3cm}
\end{center}
\caption{Same as Fig. \ref{f3} for $I=3/2$.}%
\label{f4}%
\end{figure}

In Figs.~\ref{f3} and \ref{f4} we show $|T|^{2}$ as a function of $\sqrt{s}$
for $\rho\Delta\rightarrow\rho\Delta$ both in $I=1/2$ and $I=3/2$. In order to
check the dependence of our results on the choice made for the value of the
cutoff, a $10$\% variation over the value taken in Ref.~\cite{sarkar}
($q_{max}=700$ MeV) has been considered. We shall take the corresponding
dispersion of results as an indication of their uncertainty. By comparing
Figs.~\ref{f3} and \ref{f4} it is clear that the results for $I=3/2,$ with
bound state masses ranging from $1940$ MeV to $1980$ MeV, show much less
dispersion than the results for $I=1/2,$ with bound state masses ranging from
approximately $1700$ MeV to $1800$ MeV. This has to do with the more
attractive interaction for $I=1/2$, so that the more bound the sate the more
uncertain the prediction of its mass.

We postpone the comparison with experimental $N^{\ast}$ and $\Delta^{\ast}$
resonances until a discussion of possible corrections to these results coming
from the inclusion of $\rho$ and $\Delta$ widths or from anomalous terms is
carried out.

\section{Convolution due to the $\rho$ and $\Delta$ mass distributions}

\label{s4} The strong attraction in the $I=1/2,3/2$ channels produces
$\rho\Delta$ bound states and thus with no width within the model. However,
this is not strictly true since the $\rho$ and $\Delta$ have a large width, or
equivalently a mass distribution that allows the states obtained to decay in
$\rho\Delta$ for the low mass components of the $\rho$ and $\Delta$ mass
distributions. To take this into account we follow the usual procedure
consisting in convoluting the $G$ function with the mass distributions of the
$\rho$ and $\Delta$~\cite{nagahiro} so that the $G$ function is replaced by
$\tilde{G}$ as follows
\begin{align}
\tilde{G}(s) &  =\frac{1}{N_{\rho}N_{\Delta}}\int_{m_{\Delta}-2\Gamma_{\Delta
}}^{m_{\Delta}+2\Gamma_{\Delta}}d\tilde{M}(-\frac{1}{\pi})\mathcal{I}m\frac
{1}{\tilde{M}-M_{\Delta}+i\frac{\Gamma_{1}(\tilde{M})}{2}}\nonumber\\
&  \times\int_{(m_{\rho}-2\Gamma_{\rho})^{2}}^{(m_{\rho}+2\Gamma_{\rho})^{2}%
}d\tilde{m}^{2}(-\frac{1}{\pi})\mathcal{I}m\frac{1}{\tilde{m}^{2}-m_{\rho}%
^{2}+i\tilde{m}\Gamma_{2}(\tilde{m})}G(s,\tilde{M},\tilde{m}^{2}%
)\ ,\label{Gconvolution}%
\end{align}
with
\begin{align}
N_{\rho} &  =\int_{(m_{\rho}-2\Gamma_{\rho})^{2}}^{(m_{\rho}+2\Gamma_{\rho
})^{2}}d\tilde{m}^{2}(-\frac{1}{\pi})\mathcal{I}m\frac{1}{\tilde{m}%
^{2}-m_{\rho}^{2}+i\tilde{m}\Gamma_{2}(\tilde{m})}\\
N_{\Delta} &  =\int_{m_{\Delta}-2\Gamma_{\Delta}}^{m_{\Delta}+2\Gamma_{\Delta
}}d\tilde{M}(-\frac{1}{\pi})\mathcal{I}m\frac{1}{\tilde{M}-M_{\Delta}%
+i\frac{\Gamma_{1}(\tilde{M})}{2}}\,,\nonumber\label{Norm}%
\end{align}
where
\begin{align}
\Gamma_{1}(\tilde{M}) &  =\Gamma_{\Delta}\left(  \frac{\lambda^{1/2}(\tilde
{M}^{2},M_{N}^{2},m_{\pi}^{2})2M_{\Delta}}{\lambda^{1/2}(M_{\Delta}^{2}%
,M_{N}^{2},m_{\pi}^{2})2\tilde{M}}\right)  ^{3}\theta(\tilde{M}-M_{N}-m_{\pi
})\\
\Gamma_{2}(\tilde{m}) &  =\Gamma_{\rho}(\frac{\tilde{m}^{2}-4m_{\pi}^{2}%
}{m_{\rho}^{2}-4m_{\pi}^{2}})^{3/2}\theta(\tilde{m}-2m_{\pi}%
)\nonumber\label{gamma}%
\end{align}
with $\lambda(x,y,z)=x^{2}+y^{2}+z^{2}-2xy-2xz-2yz$, $\Gamma_{\Delta}=120$ MeV
and $\Gamma_{\rho}=150$ MeV (for $\Gamma_{2}(\tilde{m})$ we have taken the
$\rho$ width for the decay into two pions in $P$-wave). The use of $\tilde{G}$
in Eq.~(\ref{15}) provides a width to the states obtained.

\section{Results with $\rho$ and $\Delta$ widths}

\label{s5} 
\begin{figure}[tb]
\begin{center}
\vspace*{-2cm} \epsfig{file=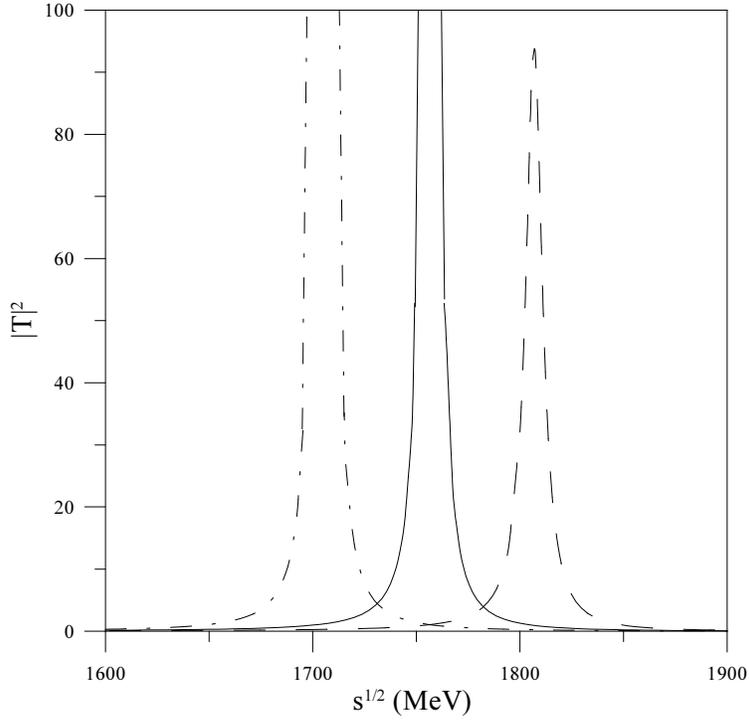, width=10cm} \vspace*{-3cm}
\end{center}
\caption{$|T|^{2}$ for $\rho\Delta\rightarrow\rho\Delta$ in the $I=1/2$
channel for several values of the cutoff $q_{max}$ including $\rho$ and
$\Delta$ mass distributions: solid line $q_{max}=770$ MeV, dashed line
$q_{max}=700$ MeV, dashed-dotted line $q_{max}=630$ MeV.}%
\label{f5}%
\end{figure}

\begin{figure}[tb]
\begin{center}
\vspace*{-2cm} \epsfig{file=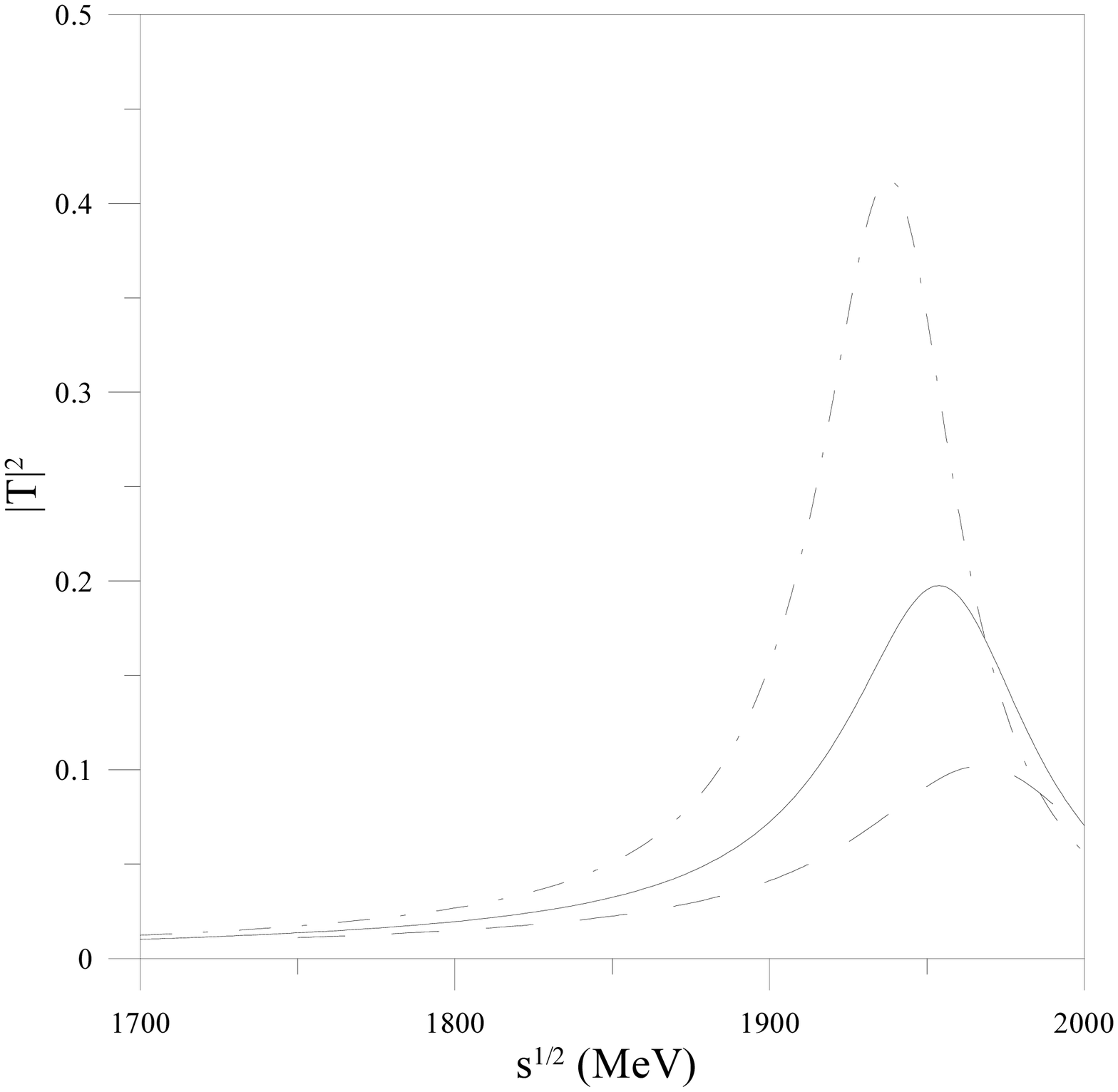, width=10cm} \vspace*{-3cm}
\end{center}
\caption{Same as Fig. \ref{f5} for $I=3/2$.}%
\label{f6}%
\end{figure}

In Figs.~\ref{f5} and \ref{f6} we show the results taking into
account the widths of the $\rho$ and $\Delta$ as discussed in the previous
section. We can see that the states develop a width leading to more realistic
results. Yet, this is not the whole width of the states since the $\pi N$
decay channels have not been included in the approach. As discussed in
Ref.~\cite{sarkar} these channels are supposed to play a minor role as
building blocks of the resonance since they are far apart in energy from the
masses of the states obtained. However, given the fact that there is more
phase space for the decay into these channels, they can have some contribution
to the total width. One should note though that the inclusion of the width
hardly changes the position of the peaks.

\section{Contribution from anomalous terms}

\label{s6} \begin{figure}[tb]
\begin{center}
\epsfig{file=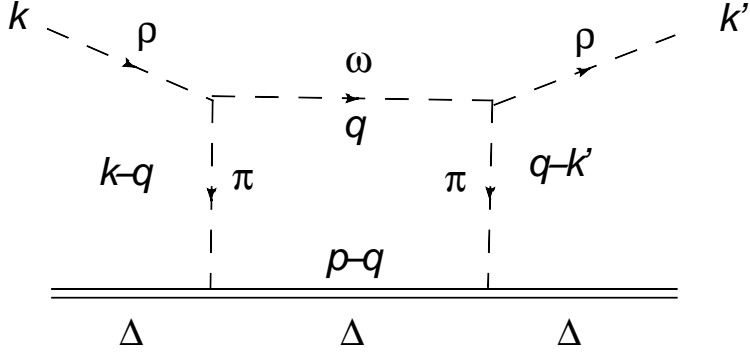, width=10cm}
\end{center}
\caption{$\rho\Delta\to\rho\Delta$ term involving $\omega\Delta$ intermediate
states through anomalous $\rho\to\pi\omega$ couplings.}%
\label{f7}%
\end{figure}

In the previous sections we mentioned that we do not have contributions from
either $\rho\Delta\rightarrow\omega\Delta$ or $\omega\Delta\rightarrow
\omega\Delta$ transitions within the vector exchange approach of our model.
Yet it is possible to have $\omega\Delta$ intermediate states through
anomalous terms involving the $\rho\rightarrow\omega\pi$ transition. The
contribution to $\rho\Delta\rightarrow\rho\Delta$ from intermediate
$\omega\Delta$ states is studied through the diagram depicted in
Fig.~\ref{f7}. The anomalous coupling $\rho\rightarrow\omega\pi$ is considered
using the same normalization of Ref.~\cite{nagahiro},
\begin{equation}
\mathcal{L}_{VVP}=\frac{G^{\prime}}{\sqrt{2}}\epsilon^{\mu\nu\alpha\beta
}\langle\partial_{\mu}V_{\nu}\partial_{\alpha}V_{\beta}P\rangle\label{18}%
\end{equation}
where $G^{\prime}=3g^{\prime}{}^{2}/4\pi^{2}f$, $g^{\prime}=-G_{V}M_{\rho
}/\sqrt{2}f^{2}$, $G_{V}\approx55$ MeV. For $\rho^{+}(k)\rightarrow
\omega(q)\pi^{+}$ we get the coupling
\begin{equation}
-it=iG^{\prime}\epsilon^{\mu\nu\alpha\beta}q_{\mu}k_{\alpha}\epsilon_{\nu
}(\omega)\epsilon_{\beta}(\rho^{+})\label{19}%
\end{equation}
and the same contribution for $\rho^{0}\rightarrow\omega\pi^{0}$.

However, it must be taken into account that the three momenta of the external
vector mesons has been neglected and, thus, $k_{\alpha}\to k_{0}$, forcing
$\mu,\nu$, and $\beta$ to be spatial components. Hence,
\begin{equation}
-it=-iM_{\rho}G^{\prime}\epsilon_{ijl}q^{i}\epsilon^{j}(\omega)\epsilon
^{l}(\rho^{+})\,. \label{20}%
\end{equation}

Note that in the diagram of Fig.~\ref{f7} one could also have a $N$ in the
intermediate state. We show next that the contribution of the diagram of
Fig.~\ref{f7} is pretty small. Therefore we shall neglect it as well as the
one with an intermediate nucleon.

To evaluate the diagram of Fig.~\ref{f7} one needs the $\pi\Delta\Delta$
coupling, which have been taken from~\cite{tejedor},
\begin{equation}
\mathcal{L}_{\Delta\Delta\pi}=-\frac{f_{\Delta}}{m_{\pi}}\Psi_{\Delta}%
^{+}S_{\Delta,i}(\partial_{i}\phi^{\lambda})T_{\Delta}^{\lambda}\Psi_{\Delta}
\label{20b}%
\end{equation}
with $f_{\Delta}=0.802$, where $\vec{S}_{\Delta}(\vec{T}_{\Delta})$ is the
spin(isospin) operator for the $\Delta$, $\vec{S}_{\Delta}^{2}=S(S+1)$ and
$\vec{T}_{\Delta}^{2}=T(T+1))$.

By taking the usual isospin convention $|\rho^{+}\rangle=-|1,+1\rangle$, one
has
\begin{align}
|\Delta\rho;3/2,+3/2\rangle &  =\sqrt{\frac{3}{5}}\Delta^{++}\rho^{0}+
\sqrt{\frac{2}{5}}\Delta^{+}\rho^{+}\\
|\Delta\rho;1/2,+1/2\rangle &  =\sqrt{\frac{1}{2}}\Delta^{++}\rho^{-}-
\sqrt{\frac{1}{3}}\Delta^{+}\rho^{0}-\sqrt{\frac{1}{6}}\Delta^{0}\rho^{+}
\nonumber\label{21}%
\end{align}

The only contribution obtained corresponds to isospin $I=3/2$ $\rho\Delta$
states, since $\omega\Delta$ only couples to this isospin. In doing so one
obtains
\begin{equation}
\langle\Delta\rho,I=3/2|(\vec{T}_{\Delta}\vec{\phi})^{2}|\Delta\rho
,I=3/2\rangle=\frac{15}{4}\label{22}%
\end{equation}
by means of which the contribution of the diagram of Fig.~\ref{f7} to the
$\rho\Delta\rightarrow\rho\Delta$ transition is given by
\begin{align}
-it &  =\frac{15}{4}M_{\rho}^{2}(G^{\prime})^{2}\left(  \frac{f_{\Delta}%
}{m_{\pi}}\right)  ^{2}\epsilon_{ijk}\epsilon_{i^{\prime}j^{\prime}k^{\prime}%
}(-i)(-i)\int\frac{d^{4}q}{(2\pi)^{4}}q^{i}\epsilon^{j}(\omega)\epsilon
^{k}(\rho)q^{i^{\prime}}\epsilon^{j^{\prime}}(\omega)\epsilon^{k^{\prime}%
}(\rho)\times\nonumber\\
&  \times\frac{i}{(k-q)^{2}-m_{\pi}^{2}+i\epsilon}\,\,\frac{i}{(q-k^{\prime
})^{2}-m_{\pi}^{2}+i\epsilon}\left(  \vec{S}_{\Delta}\cdot\vec{q}\right)
\left(  \vec{S}_{\Delta}\dot{(}-\vec{q})\right)  \times\nonumber\\
&  \times\frac{i}{q^{2}-M_{\omega}^{2}+i\epsilon}\,\,\frac{i}{P-q-E_{\Delta
}(q)+i\epsilon}\label{22b}%
\end{align}

By summing over the polarizations, setting $\vec{k}$, $\vec{k}^{\prime}$ equal
to zero, taking into account that
\begin{equation}
\int\frac{d^{3}q}{(2\pi)^{3}}f(\vec{q}^{\,2})q_{i}q_{j}q_{l}q_{m}=\frac{1}%
{15}\int\frac{d^{3}q}{(2\pi)^{3}}f(\vec{q}^{\,2})\vec{q}^{\,\,4}(\delta
_{ij}\delta_{lm}+\delta_{il}\delta_{jm}+\delta_{im}\delta_{jl})\,,\label{23}%
\end{equation}
and ignoring the tensor part $(S_{\Delta_{i}}S_{\Delta_{j}}-\frac{1}{3}\vec
{S}_{\Delta}^{2}\delta_{ij})\epsilon_{i}\epsilon_{j}^{\prime}$ versus the
dominant scalar part $\frac{5}{3}\vec{\epsilon}\,\vec{\epsilon}^{\text{
}\prime}\vec{S}_{\Delta}^{2}=\frac{25}{4}\vec{\epsilon}\,\vec{\epsilon
}^{\text{ }\prime}$ (we estimate the contribution of the tensor part to be of
the order of 10\% of the scalar one), one arrives to
\begin{align}
-it &  =\frac{25}{8}\vec{\epsilon}\,\vec{\epsilon}^{\text{ }\prime}\left(
M_{\rho}G^{\prime}\frac{f_{\Delta}}{m_{\pi}}\right)  ^{2}\int\frac{d^{4}%
q}{(2\pi)^{4}}\vec{q}^{\,\,4}\frac{1}{(k^{\prime}-q)^{2}-m_{\pi}^{2}%
+i\epsilon}\times\\
&  \times\frac{1}{(q-k)^{2}-m_{\pi}^{2}+i\epsilon}\,\,\frac{1}{q^{2}%
-M_{\omega}^{2}+i\epsilon}\,\,\frac{1}{(P^{0}-q^{0})-E_{\Delta}(q)+i\epsilon
}\nonumber%
\end{align}
The $q^{0}$ integration is now performed summing the residues of the poles of
the propagator and one finally finds
\begin{align}
t_{\rho\Delta\rightarrow\rho\Delta}^{(anomalous)} &  =\frac{25}{8}%
\vec{\epsilon}\,\vec{\epsilon}^{\text{ }\prime}\left(  M_{\rho}G^{\prime}%
\frac{f_{\Delta}}{m_{\pi}}\right)  ^{2}\int\frac{d^{3}q}{(2\pi)^{3}}\vec
{q}^{\,\,4}\frac{1}{2\omega}\frac{1}{(P^{0}-\omega)-E_{\Delta}+i\epsilon
}\times\label{25}\\
&  \times\left(  \frac{1}{\omega-k^{0}+\omega_{\pi}}\right)  ^{2}\left(
\frac{1}{\omega+k^{0}+\omega_{\pi}}\right)  ^{2}\left(  \frac{1}{P^{0}%
-k^{0}-\omega_{\pi}-E_{\Delta}}\right)  ^{2}\frac{1}{(2\omega_{\pi})^{3}%
}4\times\nonumber\\
&  \times\left\{  2\omega_{\pi}^{5}+4(E_{\Delta}+k^{0}+2\omega-P^{0}%
)\omega_{\pi}^{4}+2[E_{\Delta}^{2}+2k^{0}E_{\Delta}+6\omega E_{\Delta}\right.
\nonumber\\
&  +(k^{0})^{2}+6\omega^{2}+(P^{0})^{2}+4k^{0}\omega-2(E_{\Delta}%
+k^{0}+3\omega)P^{0}]\omega_{\pi}^{3}+\nonumber\\
&  +\omega(5E_{\Delta}+7k^{0}+8\omega-5P^{0})(E_{\Delta}+\omega-P^{0}%
)\omega_{\pi}^{2}+2\omega(E_{\Delta}+\omega-P^{0})\omega_{\pi}\times
\nonumber\\
&  \times\lbrack-(k^{0})^{2}+2\omega k^{0}+\omega(2E_{\Delta}+\omega
-2P^{0})]+\omega(E_{\Delta}+\omega-P^{0})\times\nonumber\\
&  \times(E_{\Delta}+k^{0}-P^{0})(\omega^{2}-(k^{0})^{2})\left.  {}\right\}
FF(\vec{q}\,)^{2}\nonumber%
\end{align}
In Eq.~(\ref{25}) a form factor
\begin{equation}
FF(\vec{q})=\frac{\Lambda^{2}}{\Lambda^{2}-\vec{q}^{\,2}}\label{26}%
\end{equation}
has been included for each $\Delta\Delta\pi$ vertex, with $\Lambda=1$ GeV, as
customary in the Yukawa coupling of pions and baryons.

\begin{figure}[tb]
\begin{center}
\vspace*{-2cm} \epsfig{file=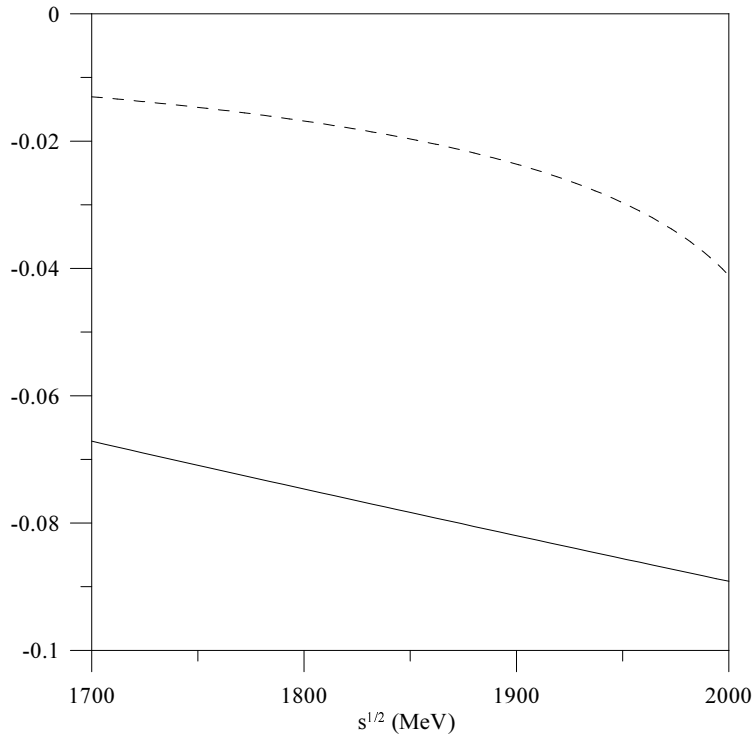, width=10cm} \vspace*{-3cm}
\end{center}
\caption{$t_{\rho\Delta\rightarrow\rho\Delta}^{(anomalous)}$ (dashed lines)
compared to $V$ (solid lines) as a function of $\sqrt{s}$ for $q_{max}=770$
MeV/c.}%
\label{f8}%
\end{figure}

In Fig.~\ref{f8} we show the results for $t_{\rho\Delta\rightarrow\rho\Delta
}^{(anomalous)}$ compared to $V$ for the same channel for the exchange of
vector mesons. As can be seen, the contribution of the anomalous term is
reasonably smaller than the dominant one of vector meson exchange and, hence,
neglecting it, as we have done in the former section, is a good approximation.
Note that this is not the only source of smaller contributions to the
potential. We have also ignored terms with $\pi N$ and $\pi\Delta$ in the
intermediate states. Actually, in Ref.~\cite{raquel}, the equivalent terms
with $\pi\pi$ intermediate states in the $\rho\rho$ scattering are found to
have a relatively small real part compared to the contribution from the
dominant $\rho$ exchange terms. Furthermore, they have opposite sign to the
anomalous contributions, leading to additional cancellations of these small
terms that we also expect here.

\section{Comparison to experimental states}

\label{s7} To compare our dynamically generated meson-baryon bound states
(DGBS) with experimental resonances we should keep in mind that a calculated
DGBS mass will only fit precisely an experimental mass if the corresponding
resonance has either i) a very dominant meson-baryon $(4q1\overline{q})$
component or ii) weakly coupled $3q$ and meson-baryon components giving rise
separately to bound states of the same mass.\ 

Option i) seems to be approximately at work at least for some of our
degenerate $I=3/2,$ $J^{P}=1/2^{-},3/2^{-},5/2^{-}$ states, with a mass
between $1940$ MeV ($q_{max}=770$ MeV) and $1980$ MeV ($q_{max}=630$ MeV),
which can be respectively assigned to $\Delta(1900)S_{31}(\ast\ast),$
$\Delta(1940)D_{33}(\ast)$ and $\Delta(1930)D_{35}(\ast\ast\ast)$ from the
Particle Data Group (PDG) Review \cite{PDG08}. Indeed the mass of these three
resonances can not be reproduced by $3q$ models based on two-quark
interactions which predict significant higher values \cite{GVV08}.
Specifically for $\Delta(1930)D_{35}$, due to quark Pauli blocking, there are
not allowed $3q$ configurations in the first energy band of negative parity
and consequently any $3q$ mass prediction is about $300$ MeV higher than the
PDG\ average mass. Then we can interpret $\Delta(1930)D_{35}$ as a $\rho
\Delta$ bound state whereas the corresponding $3q$ bound state will lye $300$
MeV above. For $J^{P}=1/2^{-},3/2^{-}$ the predicted $3q$ radial excitations
of the lowest states in the first energy band of negative parity are located
at $\sim2050$ MeV (see for instance \cite{Cap86}). If they were overpredicted,
as it is known to happen for radial excitations in the positive parity sector,
there could be forming part of $\Delta(1900)S_{31}$ and $\Delta(1940)D_{33}$
altogether with the $\rho\Delta$ bound state and option ii) would be
preferred. These considerations get additional support from some experimental
analyses \cite{Cut80,Man92} where the inclusion of $\rho\Delta$ as an
effective inelastic channel becomes essential for the extraction of the
$\Delta(1930)D_{35}.$

For $I=1/2,$ $J^{P}=1/2^{-},3/2^{-},5/2^{-}$ and $q_{max}=770$ MeV we obtain a
DGBS mass of $\sim1700$ MeV. A look to the PDG data suggests the
identification with the almost degenerate $N(1650)S_{11}(\ast\ast\ast\ast),$
$N(1700)D_{13}(\ast\ast\ast)$ and $N(1675)D_{15}(\ast\ast\ast\ast) $
respectively. However, although for these resonances $3q$ models
systematically underpredict their masses, the predicted $3q$ values differ
less than $100$ MeV from the PDG averages. This makes option ii) more reliable
as we show next. Let us centre first on $D_{15}$ for which the $3q$ mass
prediction is usually closer to data. If the ($3q-\rho\Delta)_{D_{15}} $
coupling were weak there would be two bound states close in mass, differing
very little from the $3q$ and $\rho\Delta$ bound states. As there is only one
experimental candidate in the energy region under consideration (the closest
PDG $D_{15}$ state is $N(2200)D_{15}(\ast\ast))$ we would be forced to
conclude that both states actually form the $N(1675)D_{15}.$ On the other hand
if the ($3q-\rho\Delta)_{D_{15}}$ coupling were strong, but not enough to make
the $\rho\Delta$ unbound, there would be two bound states much more separated
in mass than the previous $3q$ and $\rho\Delta$ ones (see Ref.~\cite{GVV08}).
The two new bound states should appear as two distinctive resonances what is
not confirmed experimentally. Therefore we conclude that the $N(1675)D_{15}$
may be the result of the overlapping of the $3q$ and $\rho\Delta$ bound states.

For $N(1650)S_{11}$ and $N(1700)D_{13}$ the analysis is more difficult since
there can be $3q$ configuration mixing with the low-lying $N(1535)S_{11}$ and
$N(1520)D_{13}$ and also a possible coupling to the S-wave $\rho N$ channel
that we have not taken into account. The study of the $\rho N$ channel and its
effect on the $N$ and $\Delta$ spectrum deserves special attention and will be
the subject of a future analysis.

Regarding the dependence of our results on the cutoff it should be remarked
that for any value in the considered interval $q_{max}\in$ [ 630 MeV, 770 MeV]
the only available assignment of our $I=3/2$ $\rho\Delta$ bound states to
$\Delta^{\ast}$ resonances is the one performed above. This is easy to
understand by realizing that even enlarging the $q_{max}$ interval the mass of
the closest $\Delta(J^{-})$ PDG resonances would be far above or below our
predicted masses. Moreover as the change of the prediction over the whole
interval is very modest, going from $1940$ MeV to $1980$ MeV, there can only
have a little effect on the probability of the $\rho\Delta$ component in the
assigned resonance.

For $I=1/2$ the greater sensitivity of the predicted masses to changes in the
cutoff values (from $1700$ MeV to $1800$ MeV in the considered $q_{max}$
interval) might leave room for an alternative assignment although a quick look
to the PDG catalog also locates the next $N(J^{P}=5/2^{-})$ resonance,
$N(2200)D_{15}(\ast\ast),$ very far above our predicted masses. However, a
careful look to the data values used to obtain its average mass indicates two
sets of them: one giving a mass about $1900$ MeV and another one reporting
values around $2200$ MeV. The same circumstance is repeated for $N(2090)S_{11}%
(\ast)$ and $N(2080)D_{13}(\ast\ast)$ so that the existence of two distinctive
resonances may be considered. Actually in Ref.~\cite{Cut80} two $D_{13}$
resonances with masses $2080$ MeV and $1880$ MeV were reported. Then, by
enlarging the cutoff interval, our $\rho\Delta$ bound state prediction (for
$q_{max}=530$ MeV we predict $1880$ MeV) could be pretty close in mass to the
$N(1900)(J^{P}=1/2^{-},3/2^{-},5/2^{-})$ and their assignment would be
feasible$.$ Again the presence of $3q$ models configurations close in mass
could imply a $3q-\rho\Delta$ structure. In such a case the $N(1650)S_{11},$
$N(1700)D_{13}$ and $N(1675)D_{15}$ should consistently not contain a
$\rho\Delta$ component.

\section{Conclusions}

\label{s8} We have performed a study of the $\rho\Delta$ interaction within
the framework of the hidden gauge formalism for vector mesons and a unitary
approach via the use of the Bethe Salpeter equation. We find that the
interaction potential for the $\rho$ and $\Delta$ is attractive in $I=1/2$ and
$I=3/2$ and repulsive in $I=5/2$ channels. Then, we found bound states of
$\rho\Delta$ in the $I=1/2,3/2$ channels and no bound states in the $I=5/2$
channel. It is interesting to observe that, even if the $\rho\Delta$ structure
allows for $I=5/2$, the dynamics of the problem precludes the formation of
bound states. Note that an $I=5/2$ bound state would be exotic since a three
quark structure does not allow such an isospin to appear.

We also find the $\rho\Delta$ interaction with $I=1/2$ to be stronger than
with $I=3/2$, leading to $N^{\ast}$ states more bound that the $\Delta^{\ast}$
ones. The other dynamical feature of the model is the spin degeneracy in
$J^{P}=1/2^{-},3/2^{-},5/2^{-}$ of both the $N^{\ast}$ and $\Delta^{\ast}$
states. This is a consequence of the approximations done, corresponding to the
$\rho\Delta$ interaction in $S-$ wave, neglecting the three momentum of the
vector mesons. As much as the approximations are sensible, we expect the
predictions on the degeneracy to be realistic.

When it comes to compare our results with existing resonances, we find for
$\Delta^{\ast}$ a good quantitative agreement with $\Delta(1900)S_{31}%
(\ast\ast),$ $\Delta(1940)D_{33}(\ast)$ and $\Delta(1930)D_{35}(\ast\ast
\ast).$ The small sensitivity of the predicted masses to changes in the cutoff
parameter used in our calculation added to the lack of alternative
identifications with existing data takes us to unambiguously assign our
$\rho\Delta$ bound states to these resonances. For $\Delta(1930)D_{35}$ the
much larger mass predicted by $3q$ models makes clear that it contains almost
exclusively a $\rho\Delta$ component. For $\Delta(1900)S_{31}$ and
$\Delta(1940)D_{33}$ the presence of a $3q$ radial excitation not far above in
energy could leave room for a significant probability of this component.

Concerning $N^{\ast}$ we find again a good quantitative agreement of our
predicted masses with $N(1650)S_{11}(\ast\ast\ast\ast),$ $N(1700)D_{13}%
(\ast\ast\ast)$ and $N(1675)D_{15}(\ast\ast\ast\ast).$ The analysis of $3q$
models predictions makes clear that with this assignment each of these
experimental $N^{\ast}$ resonances would come out from the overlapping of two
different states containing both $3q$ as well as $\rho\Delta$ components.
However the close mass of these two states and the possible existence of
additional components in them would make difficult their experimental
disentanglement. Alternatively another assignment is possible. The larger
sensitivity of the predicted masses to changes in the cutoff could make
feasible in this case the assignment to non-cataloged $N(1900)(J^{P}%
=1/2^{-},3/2^{-},5/2^{-})$ resonances which would be hidden in the cataloged
$N(2200)D_{15}(\ast\ast),$ $N(2090)S_{11}(\ast)$ and $N(2080)D_{13}(\ast
\ast).$

Therefore we conclude that the $\Delta(1930)D_{35}(\ast\ast\ast)$ can be
interpreted as $\rho\Delta$ bound state and that the $\Delta(1900)S_{31}%
(\ast\ast)$ and $\Delta(1940)D_{33}(\ast)$ contain a significant probability
of the $\rho\Delta$ component providing an explanation to the elusive
description of these resonances by $3q$ models. In the $I=1/2$ sector although
an assignment of our $\rho\Delta$ bound states to $N(1650)S_{11}(\ast\ast
\ast\ast),$ $N(1700)D_{13}(\ast\ast\ast)$ and $N(1675)D_{15}(\ast\ast\ast
\ast)$ can be done it would be also possible to associate them to non
cataloged $N^{\ast}$ states around 1900 MeV. Certainly a refinement of our
results, from a more complete calculation incorporating $3q$ altogether with
meson-baryon components in a consistent manner, could be more predictive in
this case. Nevertheless the detailed analysis we have performed makes clear
that an additional experimental input is needed anyhow to definitely settle
the point. We encourage an experimental effort along this line.

\section*{Acknowledgments}

This work is partly supported by DGICYT contract number FIS2006-03438, by
Spanish MCyT and EU FEDER under Contract No. FPA2007-65748 and by Spanish
Consolider Ingenio 2010 Program CPAN (CSD2007-00042). This research is part of
the EU Integrated Infrastructure Initiative Hadron Physics Project under
contract number RII3-CT-2004-506078.

\end{document}